\documentstyle[prl, aps, amssymb
]{revtex}
\begin{document}
\draft
\wideabs{
\title{$^{17}$O NMR studies of a triangular-lattice superconductor  
Na$_{x}$CoO$_{2}\cdot y_{}$H$_{2}$O
}
\author{F.L. Ning and T. Imai}
\address{Department of Physics and Astronomy,
McMaster University, Hamilton, ON L8S 4M1, CANADA}
\date{\today}
\maketitle
\begin{abstract}
We report the first $^{17}$O NMR studies of a triangular lattice 
superconductor Na$_{1/3}$CoO$_{2}\cdot \frac{4}{3}_{}$H$_{2}$O and 
the host material Na$_{x}$CoO$_{2}$ ($x=$0.35 and 0.72).  Knight shift 
measurements reveal that p-d hybridization induces sizable spin polarization 
in the O triangular lattice layers.  Water intercalation makes 
CoO$_{2}$ planes homogeneous, 
and enhances low frequency spin fluctuations  near $T_{c}=4.5$K
at some finite wave vectors different from both the ferromagnetic and ``120 degree'' modes.
\end{abstract}
}
The recent discovery of superconductivity in hydrated 
Na$_{\frac{1}{3}}$CoO$_{2} \cdot {\frac{4}{3}}$H$_{2}$O  
($T_{c}\sim4.5 K$) has generated considerable excitement 
\cite{Takada,Chou,Foo,Sakurai,Fujimoto,Alloul,Ning}.  It appears that the symmetry 
of the superconducting energy gap is 
unconventional\cite{Fujimoto}, and this material {\it may} be the first 
realization of Anderson's R.V.B. superconductor in a 
{\it triangular lattice}.  Five electrons in the low-spin state of $t_{2g}$ orbitals of Co$^{4+}$ ions
have $S=\frac{1}{2}$ in the CoO$_{2}$ triangular lattice layers.  Each Na$^{+}$ ion donates one 
electron, and fills up all $t_{2g}$ orbitals to make  $S=0$.
Accordingly, one may view the unhydrated Na$_{x}$CoO$_{2}$ as a spin $S=\frac{1}{2}$ triangular 
lattice diluted dynamically with  $S=0$ with probability $x$.  

While the underlying physics of Na$_{x}$CoO$_{2}$ resembles that 
in high $T_{c}$ cuprates, there are important 
distinctions too.  First, one needs to intercalate water molecules to transform the Fermi liquid 
phase Na$_{\frac{1}{3}}$CoO$_{2}$ into superconducting 
Na$_{\frac{1}{3}}$CoO$_{2} \cdot {\frac{4}{3}}$H$_{2}$O.  The roles 
played by water molecules are largely unknown.  Second, the O sites are not located 
within the Co triangular lattice layers.  Instead, two sheets 
of O triangular lattice sandwich the Co triangular lattice (see 
Fig.1(a)).  A natural question to ask is whether or not p-d hybridization
between O and Co layers are relevant to the properties of 
Na$_{x}$CoO$_{2}$.  Third, frustration effects between spins are 
inherent to a triangular lattice geometry \cite{Hirakawa}, and may significantly 
affect the
spin physics.  In fact, antiferromagnetic triangular lattice 
systems tend to slow down to form a ``120 degree'' non-collinear structure as sketched 
in Fig.1(a) \cite{Hirakawa}.  We also recall that Co spin fluctuations in 
superconducting Na$_{\frac{1}{3}}$CoO$_{2}\cdot \frac{4}{3}$H$_{2}$O 
are as strong as in the magnetic Na$_{x}$CoO$_{2}$ host phases \cite{Ning}.  The crucial question to address is, whether the 
typical ``120 degree'' ordering is present in the CoO$_{2}$ 
layers, and whether doped carriers stabilize a more exotic quantum 
spin liquid state.  

In this {\it Letter}, we address these key questions through the first $^{17}$O NMR investigation 
of Na$_{\frac{1}{3}}$CoO$_{2}\cdot \frac{4}{3}$H$_{2}$O and 
two representative non-superconducting host phases 
Na$_{x}$CoO$_{2}$ ($x=0.35$ and $0.72$).  First, we show from 
$^{17}$O NMR 
lineshapes that water intercalation suppresses the residual tendency 
toward charge differentiation, and makes CoO$_{2}$ planes homogeneous
 in Na$_{\frac{1}{3}}$CoO$_{2}\cdot \frac{4}{3}$H$_{2}$O.  
Second, we show from NMR Knight shifts $^{17}K$ that the {\it local} 
spin polarization at O sites, transferred from Co sites through 
p-d hybridization, is comparable to or even greater than that observed 
for the in-plane O sites of high $T_{c}$ cuprates. 
Third, we demonstrate from the $^{17}$O nuclear spin-lattice relaxation 
rate $^{17}1/T_{1}$ that 
low frequency spin fluctuations of the superconducting 
compound grow at some finite wave vector(s) {\bf q} below $\sim$50 
K, and the slowing mode is neither at {\bf q}={\bf 0} (the ferromagnetic mode) 
nor  {\bf q}=($\frac{1}{3}$, $\frac{1}{3}$) (the ``120 degree'' mode).  
To clarify the nature of spin dynamics, we take advantage of the fact that 
the $^{17}$O site 
is located right above or below the center of Co triangles as shown in 
Fig.1(a).  This means that $^{17}1/T_{1}$ divided by temperature $T$,
 \begin{equation}
          ^{17}(\frac{1}{T_{1}T}) = 
          \frac{2\gamma_{n}^{2}k_{B}}{g^{2}\mu_{B}^{2}}\sum_{{\bf q}} |A({\bf q})|^{2} \frac{\chi''({\bf 
	q},\nu_{n})}{\nu_{n}},
\label{T1}
\end{equation}
(where $\chi''({\bf q},\nu_{n})$ is the dynamical electron spin 
susceptibility at the resonance frequency $\nu_{n}$) is insensitive to 
the ``120 degree mode'' of spin fluctuations,
because the hyperfine fields from the three nearest Co sites cancel 
out.  Such canceling effects can be represented 
by the hyperfine form factor $A({\bf q})=\Sigma_{j=1 \sim 3} A_{j}exp(-i{\bf 
q} \cdot {\bf r}_{j})$ at the 
O site, where $A_{j}$ is the 
hyperfine interaction between the O 
nuclear spin of our interest and three n.n. Co electron spins, and ${\bf r}_{j}$  
represents the position vector of Co sites with respect 
to the O site.  As shown in Fig.1(b), $A({\bf 
q})=0$ at the wave vector {\bf q}=($\frac{1}{3},\frac{1}{3}$) which 
corresponds to the 120 degree spin configuration in real space  presented in 
Fig.1(a).  


We annealed the Na$_{0.72}$CoO$_{2}$ ceramic sample \cite{Ning} in $^{17}$O$_{2}$ gas at 
830$^{\circ}$C for 5 days to enrich it with $^{17}$O 
isotope.  For Na$_{0.35}$CoO$_{2}$, we soaked a portion of the 
$^{17}$O enriched Na$_{0.72}$CoO$_{2}$ from the same batch 
in 1M Br 
acetonitrile for 5 days, washed, then 
dried it at 220$^{\circ}$C for several hours in flowing $^{16}$O$_{2}$ gas.  We aligned $^{17}$O enriched 
Na$_{x}$CoO$_{2}$ samples
 in Stycast 1266 in a magnetic field of 8 
Tesla. Since the magnetic susceptibility of Na$_{x}$CoO$_{2}$ is larger 
along the ab plane 
than along the c axis\cite{Chou}, the ab plane of ceramic particles align 
themselves along the magnetic field.  We carried out $^{17}$O 
NMR measurements in  Na$_{x}$CoO$_{2}$ by applying 
an 8 Tesla magnetic field in the aligned ab direction.  The superconducting phase was obtained by 
soaking some portion of $^{17}$O enriched Na$_{0.35}$CoO$_{2}$ 
ceramics in purified water.  We verified that for $^{59}$Co, the 
$\pm\frac{7}{2}$ to $\pm\frac{5}{2}$ NQR frequency
$^{59}\nu_{NQR}=12.3$MHz and $^{59}1/T_{1}T_{NQR}=18$ 
sec$^{-1}$K$^{-1}$ observed just above $T_{c}$ agree very well with earlier 
reports\cite{Fujimoto}.  We also confirmed that 
$^{59}1/T_{1}T$ measured in zero field by NQR dips below $T_{c}$ as 
shown in Fig.4(c).  We note that these 
properties are sensitive to the quality of the sample\cite{Kyoto}.  
We used unaligned ceramics of Na$_{1/3}$CoO$_{2}\cdot \frac{4}{3}_{}$H$_{2}$O for NMR measurements,
 because alignment with Stycast 1266 tends to 
reduce water content and damage superconducting properties.

In Fig. 2, we compare the $^{17}$O NMR lineshapes for three samples.  
Since $^{17}$O has a nuclear spin $I=\frac{5}{2}$,
one would expect the aligned ceramics of Na$_{0.72}$CoO$_{2}$ 
and Na$_{0.35}$CoO$_{2}$ to show 
only five resonance peaks from $m_{z}$ to $m_{z}+1$ transitions, 
if all O sites within CoO$_{2}$ 
layers are electronically equivalent.  However, the observed 
$^{17}$O NMR lines in Fig.2(a) and (b) are a superposition 
of two sets of signals with nearly axially symmetric Knight shift and 
quadrupole interaction tensors.  
Na$_{0.72}$CoO$_{2}$ has O(A) and O(B) 
sites, as marked in Fig.2(a) by two sets of arrows,
with relative integrated intensity of $\sim$0.7 and $\sim$0.3.  
We recall that $^{59}$Co NMR also shows two inequivalent sites, Co(A) and Co(B) sites with relative 
intensity of $\sim$0.7 and $\sim$0.3 due to {\it charge differentiation}.  Co(A) is the Co$^{+3}$-like 
site with $S\sim0$, while Co(B) is  Co$^{+4}$-like site with 
$S\sim\frac{1}{2}$.  Our new $^{17}$O NMR data indicate that
{\it the differentiation of the local Co electronic environments propagates to 
the adjacent O layers through p-d hybridization}. 
Na$_{0.35}$CoO$_{2}$ also exhibits two types of O sites, 
O(C) and O(D), with roughly comparable intensities.  Unlike  
Na$_{0.72}$CoO$_{2}$, however, the difference in the local electronic environment at 
O(C) and O(D) sites is mild.  This is consistent with our 
earlier finding from Co NMR studies that the tendency toward charge 
differentiation is strongly suppressed in the Fermi 
liquid phase\cite{Ning}.  We also found that the thermal motion of 
Na$^{+}$ ions above 220K results in {\it motional narrowing}, and O(C) and O(D) lines gradually merge 
into a single line (notice that the Knight shift $^{17}K(C)\sim^{17}K(D)$ 
at 295 K in Fig.3).  This suggests that O(C) and O(D) sites are 
differentiated by the occupancy of Na$^{+}$ sites nearby.  In contrast with these unhydrated 
samples, Fig.2(c) shows that the lineshape of 
the  unaligned superconducting sample Na$_{\frac{1}{3}}$CoO$_{2}\cdot 
\frac{4}{3}$H$_{2}$O is a typical powder pattern from a 
single O site, with a nearly isotropic NMR Kngiht shift $^{17}K$ 
($\sim0.11 \%$ at 140 K) and 
a nearly axially symmetric nuclear quadrupole interaction tensor.  This immediately 
leads us to conclude that {\it water intercalation completely suppresses 
the residual tendency toward charge differentiation in O layers in 
the Fermi liquid phase Na$_{\frac{1}{3}}$CoO$_{2}$}.  The 
detailed structural analysis showed that four water 
molecules surround each Na$^{+}$ ions \cite{Jorgensen}, and our results 
suggest that the Coulomb potential from the latter 
on CoO$_{2}$ planes is shielded by the former, thereby {\it smoothing 
out} the CoO$_{2}$ layers to make it superconducting.

It is convenient to employ $^{17}$O NMR Knight shifts 
$^{17}K=(^{17}f-^{17}f_{o})/^{17}f_{o}$ to make more 
quantitative comparison of charge differentiated
O sites in Na$_{x}$CoO$_{2}$.  $^{17}f$ is the observed $m_{z}=-\frac{1}{2}$ to $+\frac{1}{2}$ 
central peak frequency,  $^{17}f_{o}=^{17}\gamma_{n}B=46.18$MHz is 
the bare resonance frequency if there were no magnetic hyperfine 
interactions between Co electron spins and O nuclear spins 
($^{17}\gamma_{n}=5.772$ MHz/Tesla 
and $B=8$ Tesla).   Using the local spin susceptibility $\chi_{spin}^{j}$ 
at the {\it j}-th Co sites, the $^{17}$O NMR Knight shift may be written as 
\begin{equation}
          ^{17}K=^{17}K_{spin}+^{17}K_{orb}=\sum_{j}{A_{j}\chi_{spin}^{j}}+^{17}K_{orb},
\label{K}
\end{equation}
where $^{17}K_{orb}$ represents the temperature independent orbital contribution (typically $0.1\%$ or 
less\cite{Korb}).  

Fig. 3 summarizes the temperature 
dependence of $^{17}K$ at various O sites in three samples.  
Notice that $^{17}K\sim0.1\%$ in all 
cases except at O(B) sites in Na$_{0.72}$CoO$_{2}$ (the vertical scale 
for the O(B) sites is separately shown on the right hand 
side of Fig. 3).  The observed 
magnitude of $^{17}K$ is somewhat smaller than 
typical values observed for the in-plane O sites in high $T_{c}$ 
cuprates (0.1\%$\sim$0.3\% \cite{Takigawa}), but quite sizable.  In other words, 
{\it p-d hybridization introduces sizable electron spin polarization in Oxygen 
p orbitals of CoO$_{2}$ layers.  Hence we cannot ignore the roles 
played by p orbitals in the electronic properties of these materials}.
Another important point to emphasize is that $^{17}K(B)$ at O(B) sites is 
extraordinarily large, and exceeds 1 \%.  Likewise, Co(B) sites exhibit a huge NMR Knight 
shift (up to $\sim9$\%), while the Co(A) sites have nearly an order 
of magnitude smaller spin contribution\cite{Ning}.  Both $^{17}K(A)$ at 
O(A) sites and $^{17}K(B)$ show an identical 
Curie-Weiss temperature dependence as the bulk averaged magnetic susceptibility 
data and $^{59}$Co Knight shifts at Co(A) and Co(B) sites\cite{Ning}.  The solid curves are 
a Curie-Weiss fit with a constant background (presumably from the orbital 
effects $^{17}K_{orb}$),  with a common Weiss temperature 
$\theta=-29\pm3$ K.  Our present $^{17}K$ data and the relative intensities of O(A) and O(B) sites 
strongly suggest that p orbitals at O(B) sites have strong hybridization with the 
d orbitals at Co(B) sites, but O(A) sites experience only mild 
effects from Co(B) sites.  Notice that 
if each Co(B) site bonds equally with all six n.n. O 
sites, 
the intensity of O(B) sites would be factor 3 larger and account for 
$\sim0.9$ of the total integrated intensity.  Instead, our NMR results 
indicate that {\it Co(B) sites transfer their spins primarily to one of the three n.n. 
O sites in each of the O layers above and below the Co layers.}  This 
somewhat unexpected result should be taken into account in the debate 
over the magnetic interlayer couplings and the local symmetry of Co 
and O orbitals.  
For example, recent neutron scattering measurements 
\cite{Keimer} showed 
that Co-Co inter-layer exchange interactions in Na$_{0.82}$CoO$_{2}$ are comparable to 
intra-layer exchange interactions despite its layered 
structure.  There is no doubt that 
the similar Co(B)-O(B)-Na-O(B)-Co(B) exchange path is responsible for the 
unexpectedly large inter-layer coupling.  

Next, we discuss the local low frequency spin fluctuations as 
observed by the $^{17}$O NMR spin-lattice relaxation rate 
$^{17}1/T_{1}T$ (see Fig. 4).  
We found that the temperature dependence of 
$^{17}1/T_{1}T$ in Na$_{0.72}$CoO$_{2}$ satisfies the same Curie-Weiss behavior as the local {\bf 
q} = {\bf 0} susceptibility as measured by $^{17}K$.  This suggests that 
the {\it ferromagnetic} mode 
of Co spin fluctuations is enhanced within each CoO$_{2}$ layer, and 
is consistent with the recent report of a type-A antiferromagnetic 
order with in-plane ferromagnetic correlations \cite{Oxford,Keimer}.  
In contrast, $^{17}1/T_{1}T$ shows qualitatively the 
same decrease with temperature as $^{17}K$ in 
Na$_{0.35}$CoO$_{2}$.  The constant behavior of 
both $^{17}1/T_{1}T$ and $^{17}K$ below $\sim$100K is consistent 
with a Korringa law $^{17}1/T_{1}T(^{17}K)^{2}=const.$ and 
resistivity $\rho \sim 
T^{2}$ \cite{Foo}, both expected for a Fermi liquid \cite{Foo}. 

Interestingly, $^{17,59}1/T_{1}T$ and $^{17}K$ in the hydrated supercondcuting sample
Na$_{\frac{1}{3}}$CoO$_{2} \cdot {\frac{4}{3}}$H$_{2}$O 
are similar to O(D) sites in the non-superconducting  
Na$_{0.35}$CoO$_{2}$ down to $\sim$50 K.  However, $^{17}K$ remains constant 
near $T_{c}$ while both $^{17}1/T_{1}T$ and $^{59}1/T_{1}T$ 
increase toward $T_{c}=4.5$K.  The latter implies the enhancement of 
low frequency spin fluctuations before superconductivity sets in.  Aside from the 
{\it cleaner} CoO$_{2}$ planes, this is the only major difference in the local 
electronic properties between the non-superconducting 
Na$_{\frac{1}{3}}$CoO$_{2}$ and superconducting 
Na$_{\frac{1}{3}}$CoO$_{2} \cdot {\frac{4}{3}}$H$_{2}$O.  As such, it is 
natural to speculate that these enhanced spin fluctuations are responsible 
for, or at least related to, 
the mechanism of superconductivity.  The question is, where in {\bf q}-space are spin fluctuations enhanced?  Given the 
ferromagnetic nature of in-plane spin correlations in 
Na$_{x}$CoO$_{2}$ with $x=0.72$ or higher, one obvious possibility 
is near {\bf q}={\bf 0}.  However, $^{17}K$ in 
Fig.3 (as well as $^{59}K$ at $^{59}$Co sites \cite{Ning}) does not show 
enhancement near $T_{c}$, hence we can rule out the 
ferromagnetic scenario.  We note that magnetic defects and/or 
impurities easily contaminate the bulk susceptibillity data 
of Na$_{\frac{1}{3}}$CoO$_{2} \cdot {\frac{4}{3}}$H$_{2}$O to give a 
false Curie behavior near $T_{c}$.  The advantage of NMR Knight shifts 
is that defect spins only broaden the NMR lineshapes without affecting the 
temperature dependence.  

Alternately, if the frustrated 120 degree mode of spin fluctuations have a large spin-spin 
correlation length, and grow 
toward $T_{c}$ near the wave 
vector {\bf q}=($\frac{1}{3},\frac{1}{3}$), then $^{59}1/T_{1}T$ 
could grow without an increase of $^{17}K$.  However, such a 
scenario seems inconsistent with an identical increase observed for $^{17}1/T_{1}T$.
Since the hyperfine form factor satisfies 
$A(\frac{1}{3},\frac{1}{3})=0$, $^{17}1/T_{1}T$ should show little 
enhancement toward $T_{c}$.  In other words, the frustrated ``120 degree'' 
configuration seems 
irrelevant in the low frequency properties of 
Na$_{\frac{1}{3}}$CoO$_{2} \cdot {\frac{4}{3}}$H$_{2}$O near $T_{c}$ \cite{Gap}.       

To summarize the results for superconducting Na$_{\frac{1}{3}}$CoO$_{2} 
\cdot {\frac{4}{3}}$H$_{2}$O, spin fluctuations grow toward $T_{c}$ at 
some finite wave vectors, 
which are different from the ferromagnetic ${\bf q}={\bf 0}$ and 
frustrated {\bf q}=($\frac{1}{3},\frac{1}{3}$).  This may be an indication that the carrier-doped triangular lattice favors 
a quantum spin liquid state.  Given that recent ARPES measurements revealed a large Fermi surface 
in Na$_{x}$CoO$_{2}$ \cite{Hazan,Ding}, the Fermi surface geometry may 
play a crucial role in deciding the nature of spin fluctuations, perhaps 
through nesting effects for a 
large $|{\bf q}|$ value(s).  Putting all the pieces together, {\it the strongly 
correlated electron
behavior above $T_{c}$ of Na$_{\frac{1}{3}}$CoO$_{2} \cdot {\frac{4}{3}}$H$_{2}$O
 shares remarkable similarities with high $T_{c}$ 
cuprates despite the triangular lattice geometry}.    



%
%

\begin{figure}
\caption{(a) The CoO$_{2}$ layers of Na$_{x}$CoO$_{2}$.  Arrows 
represent a classically stable, frustrated ``120 degree'' spin 
configuration.  Notice that 
the hyperfine fields at $^{17}$O sites cancel out in such a 
configuration.  (b) The 
hyperfine form factor $|A({\bf q})|^{2}$ at the O sites in CoO$_{2}$ 
layers plotted in the reciprocal space.  The arrow marks 
the ${\bf q}$=($\frac{1}{3}$,$\frac{1}{3}$) point where $|A({\bf 
q})|^{2}=0$.
}
\label{Structure}
\end{figure}

\begin{figure}
\caption{(a) $^{17}$O NMR lineshape of Na$_{0.72}$CoO$_{2}$ observed with $B=8$ Tesla applied 
along the magnetically aligned ab plane.  We 
applied an inversion pulse and waited for 2 msec before recording the spin 
echo intensity to separate the O(B) 
sites with ``Fast'' relaxation times.  By taking the difference between the 
whole echo intensity (``Slow'') and the ``Fast'' O(B) contributions, 
we deduced the lineshape of O(A) sites (``Difference'').  The vertical arrows identify five 
$m_{z}$ to  $m_{z}+1$ transitions of $^{17}$O (nuclear spin 
$I=\frac{5}{2}$).  (b) $^{17}$O NMR lineshape of aligned Na$_{0.35}$CoO$_{2}$.  
(c) The powder pattern $^{17}$O NMR lineshape of
Na$_{\frac{1}{3}}$CoO$_{2} \cdot {\frac{4}{3}}$H$_{2}$O ceramics.  
The less pronounced peak structures and the high and low frequency 
tails are due to random orientations of the sample. 
}
\label{Spectra}
\end{figure}

\begin{figure}
\caption{
$^{17}K$ observed for 
various $^{17}$O sites identified in Fig.2.  The solid lines are a 
best fit to a Curie-Weiss law with a constant offset, $^{17}K(A)=4.2/(T+29)+0.024$ \%, and 
$^{17}K(B)=38/(T+29)+0.11$ \%.
}
\label{Shift}
\end{figure}

\begin{figure}
\caption{$^{17}$O and  $^{59}$Co $1/T_{1}T$ in (a) Na$_{0.72}$CoO$_{2}$, (b) Na$_{0.35}$CoO$_{2}$, and 
(c) Na$_{\frac{1}{3}}$CoO$_{2} \cdot {\frac{4}{3}}$H$_{2}$O.  In (a), 
red and blue solid curves represent a best fit to a Curie-Weiss 
behavior with the same negative Weiss temperature found for $^{17}K$ 
in Fig.3, $^{17}1/T_{1}T=29/(T+29)+0.2$ sec$^{-1}$K$^{-1}$ for O(A) 
and $^{17}1/T_{1}T=88/(T+29)+0.08$ sec$^{-1}$K$^{-1}$ for O(B).  The 
black curve is $^{59}1/T_{1}T = 241/(T+29) + 5.8 exp(-208/T)$ sec$^{-1}$K$^{-1}$, i.e. the same form of 
a Curie-Weiss term, plus an activation-type term with an energy gap 208 
K.    The $^{59}$Co NMR data in (a) and (b) are from Ning et al. [7].
}
\label{T1cap}
\end{figure}

%
%


\begin{references}
\bibitem{Korb} Our very preliminary $^{17}K$ vs. $\chi$ analysis 
showed $^{17}K_{orb} \sim $0.03\% in Na$_{1/3}$CoO$_{2}\cdot \frac{4}{3}_{}$H$_{2}$O.
\bibitem{Gap}The decrease of $^{59}1/T_{1}T$ down to $\sim$50K is faster than $^{17}1/T_{1}T$.  This may signal a pseudo gap behavior in 
the frustrated ``120 degree mode''.
\end{references}
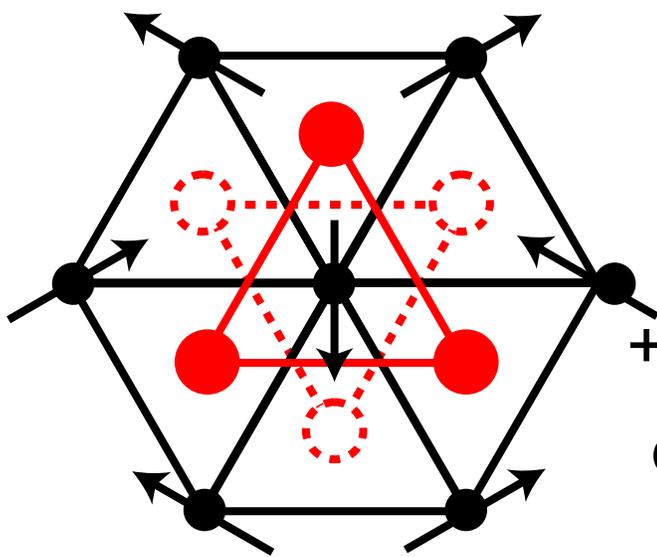
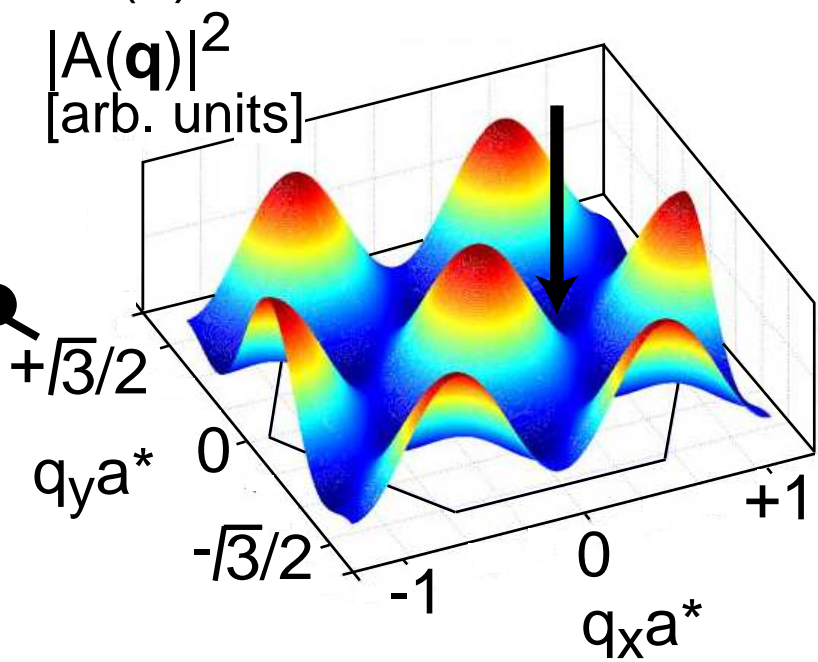
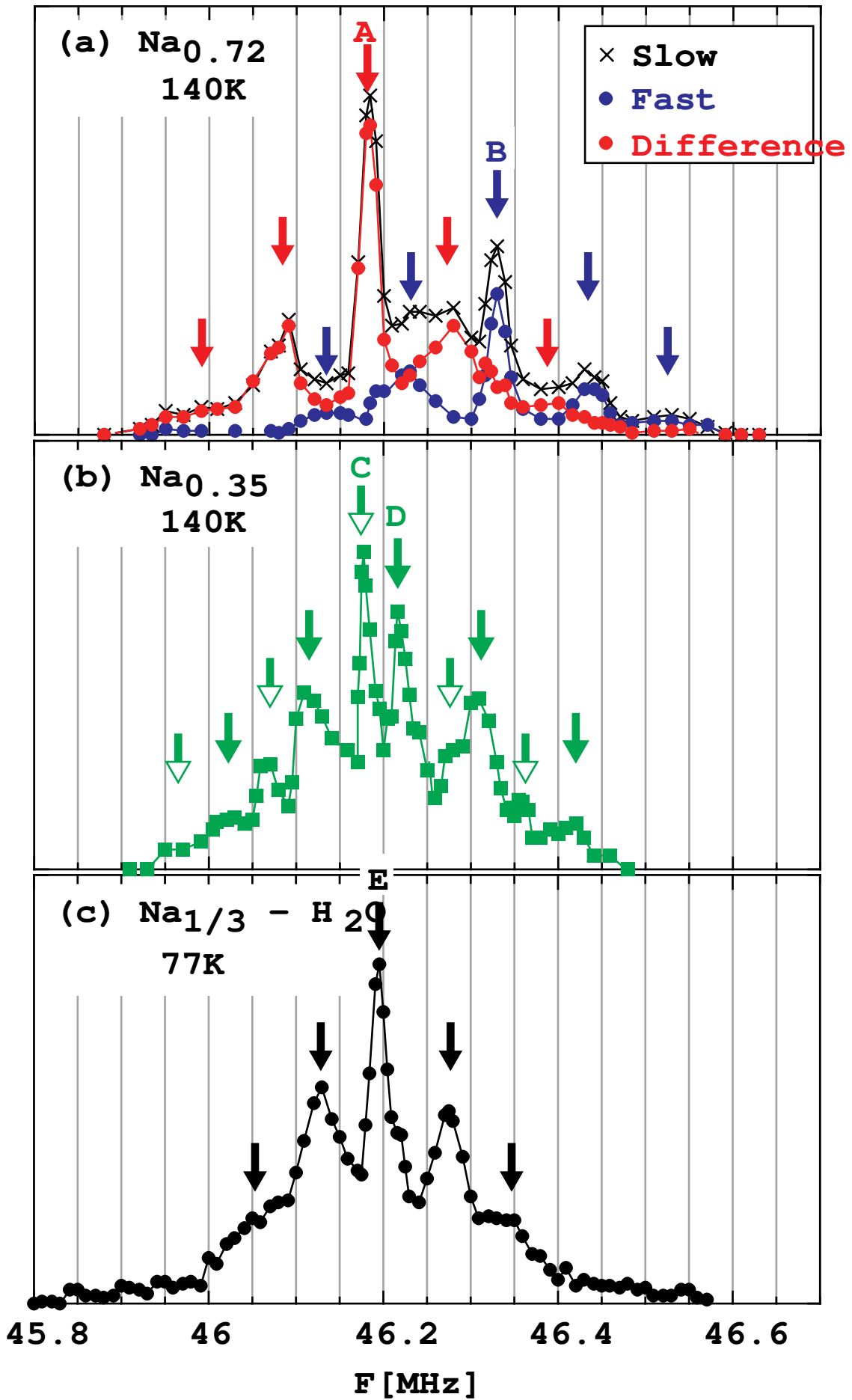
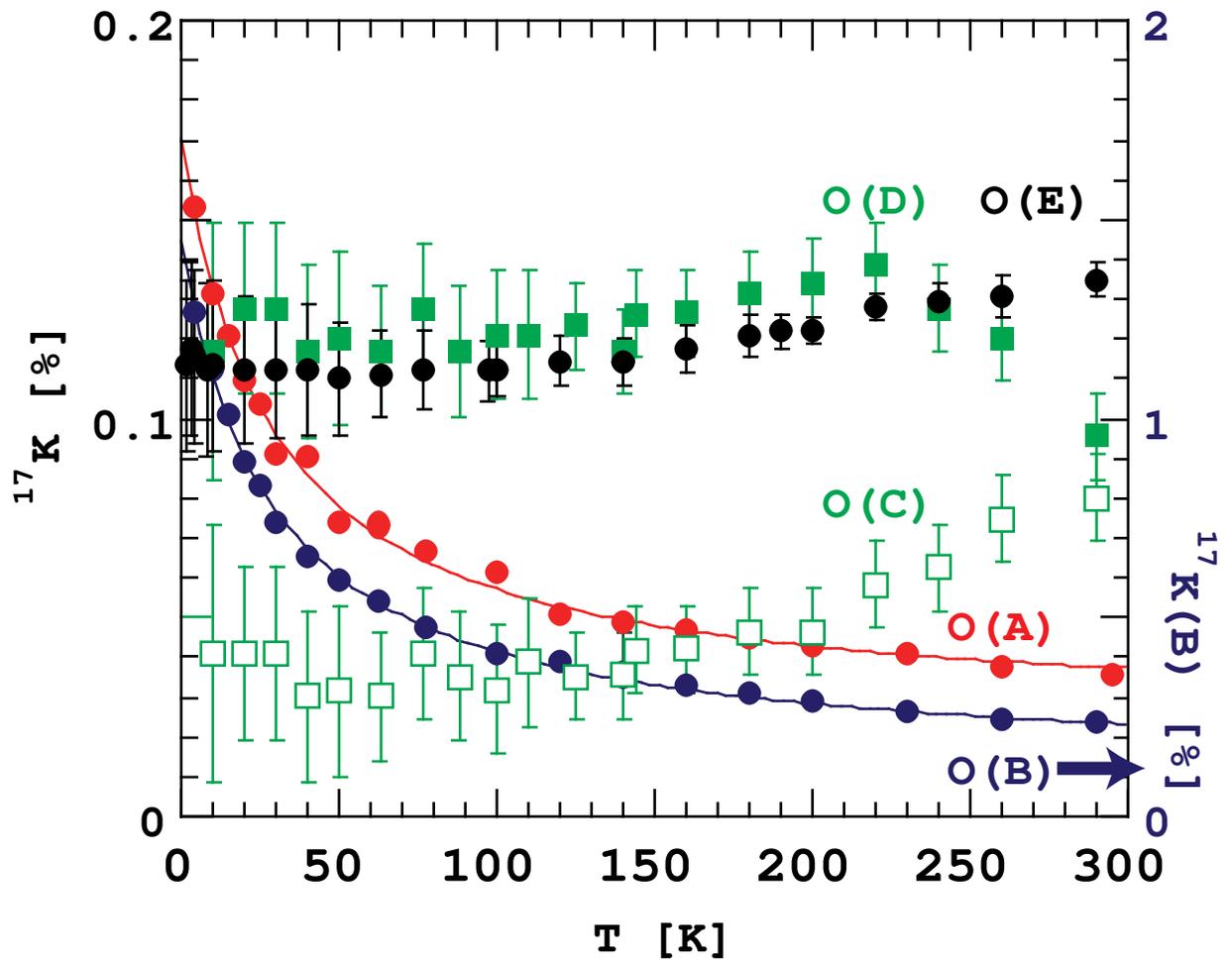
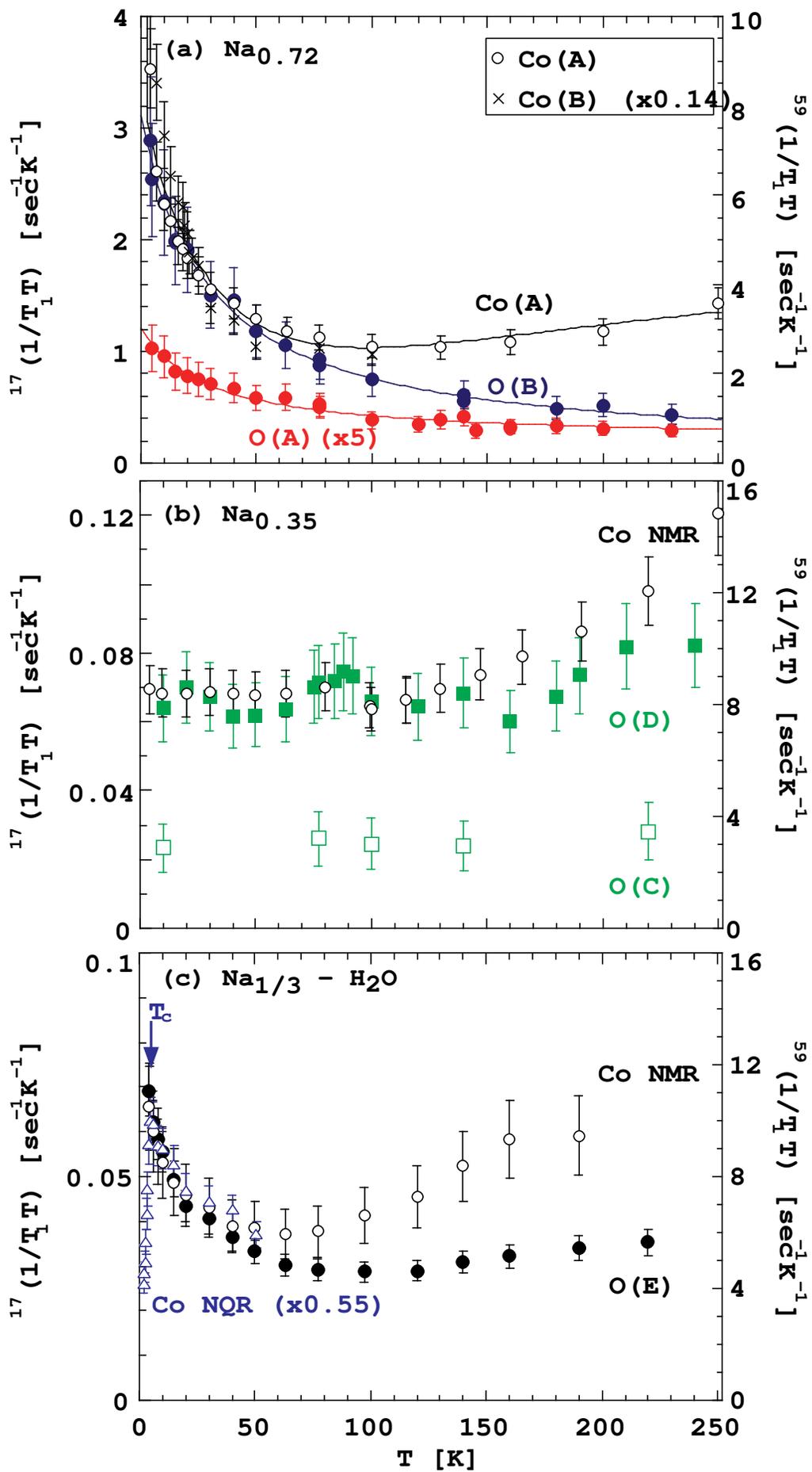
\end{document}